\documentclass[apj]{emulateapj}
\newcommand{\etal}{\mbox{~et~al.}}

\newcommand{\flux}{erg cm$^{-2}$ s$^{-1}$}
\newcommand{\xmm}{XMM--{\it Newton}}
\newcommand{\xrb}{\flux~deg$^{-2}$}
\def\deg      {{\ifmmode^\circ\else$^\circ$\fi}} 

\shorttitle{XMM-COSMOS survey II: the X-ray data and the logN-logS}
\shortauthors{N.~Cappelluti et al.}
\begin{document}
 \title{ The XMM--{\it Newton} wide-field survey in the COSMOS field II: X-ray data and the logN-logS\altaffilmark{a}}

 \author{ 
N.~Cappelluti\altaffilmark{1},
G.~Hasinger\altaffilmark{1},
M.~Brusa\altaffilmark{1},
A.~Comastri\altaffilmark{2},
G.~Zamorani\altaffilmark{2},
H.~B\"ohringer\altaffilmark{1},
H.~Brunner\altaffilmark{1},
F.~Civano\altaffilmark{2,4},
A.~Finoguenov\altaffilmark{1},
F.~Fiore\altaffilmark{5}, 
R.~Gilli\altaffilmark{2},
R.~E. Griffiths\altaffilmark{3},
V.~Mainieri\altaffilmark{1},
I.~Matute\altaffilmark{1},
T.~Miyaji\altaffilmark{3},
J.~Silverman\altaffilmark{1}
}

\altaffiltext{1}{Max Planck Institut f\"ur Extraterrestrische Physik,~D-85478 Garching, Germany}
\altaffiltext{2}{INAF-Osservatorio Astronomico di Bologna, via Ranzani 1, I-40127 Bologna, Italy}
\altaffiltext{3}{Department of Physics, Carnegie Mellon University, 5000 Forbes Avenue, Pittsburgh, PA 15213}
\altaffiltext{4}{Dipartimento di Astronomia, Universit\`a di Bologna,
    via Ranzani 1, I--40127 Bologna, Italy}
\altaffiltext{5}{INAF-Osservatorio astronomico di Roma, Via Frascati 33, I-00044 Monteporzio Catone, Italy}
\altaffiltext{a}{Based on observations
   obtained with XMM--{\it Newton,} an ESA science mission with
instruments and contributions directly funded by ESA Member States and NASA;
also based on data collected at the Canada-France-Hawaii Telescope
operated by the National Research Council of Canada,  the Centre National de
la Recherche Scientifique de France and the University of Hawaii.}
\begin{abstract}
We present the  data analysis and the X--ray source  counts for the first 
season of \xmm ~ observations in the COSMOS field.  The survey covers $\sim$2 deg$^{2}$ 
within the region of sky bounded by 
$9^h57.5^m<R.A.<10^h03.5^m$; $1^d27.5^m<DEC<2^d57.5^m$ with a total net 
integration time of 504 ks. A maximum likelihood source detection was performed in 
the 0.5--2 keV, 2--4.5 keV and 4.5--10 keV energy bands and 1390 point-like sources were 
detected in at least  one band.  Detailed Monte Carlo simulations were performed to fully   test 
the  source detection method and to derive the sky coverage to be used in the computation of the 
logN-logS relations. These relations 
have been then derived in the 0.5--2 keV, 2--10 keV and 5--10 keV energy bands, down to  flux 
limits of  7.2$\times$10$^{-16}$ erg cm$^{-2}$ s$^{-1}$, 4.0$\times$10$^{-15}$ 
erg cm$^{-2}$ s$^{-1}$ and  9.7$\times$10$^{-15}$ erg cm$^{-2}$ s$^{-1}$, respectively.
Thanks to the large number of sources 
detected in the COSMOS survey, the logN-logS curves are tightly constrained 
over a range of fluxes which were poorly covered by previous surveys, 
especially in the 2--10 and 5--10 keV bands.
 The 0.5--2 keV and  2--10 keV differential logN-logS were fitted with a 
broken power-law model which revealed   
a Euclidean slope ($\alpha\sim$2.5) at the bright end and a flatter slope ($\alpha\sim$1.5)
at faint fluxes. 
In the 5--10 keV energy band a single power-law provides an acceptable fit 
to the observed source counts with a slope 
$\alpha\sim$2.4. 
 A comparison with the results of previous surveys 
shows  good agreement in all the energy bands under investigation in the 
overlapping flux range.  
 We also notice a remarkable agreement between our logN-logS relations and the most recent model of the XRB. The slightly different normalizations observed in the source counts of COSMOS and previous surveys can be largely explained as a combination of low counting statistics and cosmic variance introduced by the large scale structure.

\end{abstract}
 \section{Introduction}
The source content of the X--ray sky has been investigated over a broad
range of fluxes and solid angles thanks to a large number of deep and
wide surveys performed in the last few years using {\it ROSAT}, {\it Chandra} and \xmm~ 
(see \citealt{bra05} for a review).
Follow--up observations unambiguously indicate that
Active Galactic Nuclei (AGN), many of which are obscured,
dominate the global energy output recorded in the cosmic
X--ray background.
The impressive amount of X--ray and multi-wavelength data obtained to date
have opened up the quantitative study
of the demography and evolution of accretion driven
Supermassive Black Holes  \citep[SMBHs;][]{myj2000,has05,ue03,laf05}.
At present the two deepest X--ray surveys, the {\it Chandra} Deep
Field North \citep[CDFN;][]{bau04} and  {\it Chandra} Deep Field South \citep[CDFS;][]{gia01},
have extended the sensitivity by about two orders of magnitude in all bands with
respect to previous surveys \citep{has93,ue99,gio00}, detecting
a large number  of faint X--ray sources.
However, deep  pencil beam surveys are limited by the area which can
be covered to very faint fluxes (typically of the order of 0.1 deg$^2$) and
suffer from significant field to field variance.
In order to cope with such  limitations, shallower surveys over  larger
areas have been undertaken in the last few years with both {\it Chandra} 
(e.g. the 9 deg$^{2}$ Bootes survey \citep{mu05}, the Extended Groth strip EGS \citep{nan05},
  the Extended {\it Chandra} Deep Field South E-CDFS, \citep{leh05,vir06} and the Champ \citep{gre04,kim04})
and \xmm ~(e.g. the HELLAS2XMM survey \citep{fio03}, the \xmm ~ BSS \citep{dc04} and the  ELAIS S1 survey \citep{puc06} ).

In this context the \xmm~ wide field
survey in the COSMOS field \citep{scoville}, hereinafter XMM--COSMOS \citep{has06}, has been conceived and designed to maximize
the sensitivity and survey area product, and  is expected to provide
a major step forward toward a complete characterization of
the physical properties of X--ray emitting SMBHs.
A contiguous area of about 2 deg$^{2}$ will be covered by
25 individual pointings, repeated twice, for a total exposure time of
about 60 ksec in each field.
 In the first observing run obtained in AO3 (phase A),   the pointings were  disposed 
 on a 5x5 grid  with the aimpoints shifted of   15' each other, 
 so as to produce a contiguous pattern of coverage.
In the second run,
to be observed in AO4 (phase B), the same pattern will be repeated
with each pointing shifted by 1' with respect to phase A.
The above described approach ensures a uniform and relatively deep coverage
of more than 1 deg$^{2}$ in the central part of the field.
When completed, XMM--COSMOS will provide an unprecedentedly large sample of
about 2000 X--ray sources with  full multi-wavelength
photometric coverage and a high level of
spectroscopic completeness.
As a consequence, the XMM--COSMOS survey is particularly well
suited to address  AGN evolution in the context
of the Large Scale Structure in which they reside.
More specifically, it will be possible to investigate
if obscured AGN are  biased tracers of the
cosmic web  and whether their space density rises
in the proximity of galaxy clusters
 \citep{hen91,max01,gil03,jo03,ya03,cap05,rud05,myj06,ya06,caps06}.\\
The X--ray  reduction of phase A data along with a
detailed analysis of the source counts in different energy bands
are presented in this paper which is organized
as follows. In Section 2 the data reduction procedure and the
relative astrometric corrections are described. In Section 3
the source detection algorithms and technique are discussed. Monte Carlo simulations 
 are presented in Section 4. The logN--logS  relations and the analysis of the contribution 
 of the XMM-COSMOS sources to the X-ray background are discussed in Section 5.
 The study of sample variance is presented in Section 6 and a summary of the work is
reported in Section 7.
The strategy and the log of the observations of XMM--COSMOS are presented by \cite{has06}, the optical identifications of X-ray sources by \cite{brus06}, the analysis of groups and clusters by \cite{fin06}, the spectral analysis of a subsample of bright sources by \cite{mai06} and the clustering of X-ray extragalactic sources by \cite{myj06}.
Throughout the paper the concordance WMAP $\Lambda$CDM
cosmology \citep{sper} is adopted
with H$_{0}$=70 km s$^{-1}$ Mpc$^{-1}$,
$\Omega_{\Lambda}$=0.7 and $\Omega_{m}$=0.3
\section{EPIC Data cleaning}
The EPIC data were processed using the \xmm~ Standard Analysis System 
(hereinafter SAS) version 6.5.0.  The Observational Data Files (ODF, "raw
data") of each of the 25 observations, were calibrated using the SAS tools \textit{epchain} and \textit{emchain}  with the most recent calibration data files. Events in bad columns, bad pixels and close to the chip gaps were excluded. \\ Both the EPIC PN and MOS event files were  searched for high particle background intervals. The distribution of the background counts binned in  100 s intervals was obtained in the 12--14 keV band for  the PN and in the 10--12 keV band for the MOS, which are dominated by particle background,  and then fitted with a gaussian model. All  time intervals with background count rate higher than 3$\sigma$ above the average best fit value were discarded. 
\begin{figure}[ht]
\center
\includegraphics[width=0.5\textwidth]{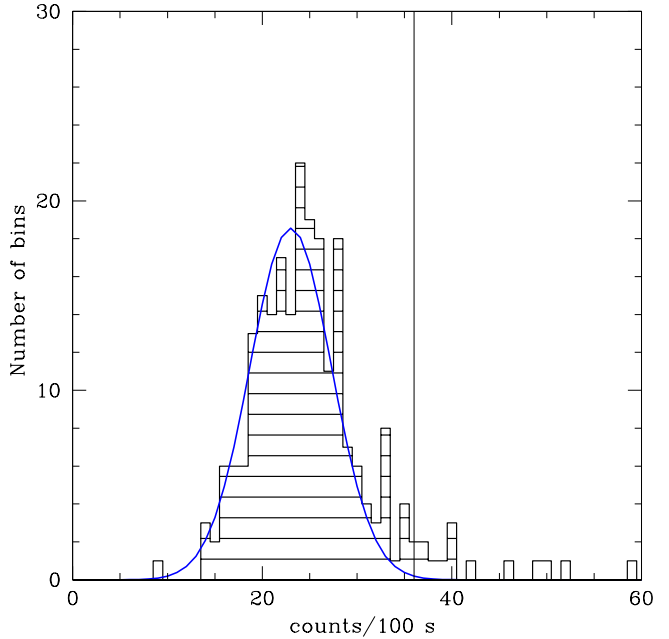}
\caption{ The background counts distribution in the PN observation of
    Field 6. The solid line represents the best  gaussian fit to the
    distribution. The continuous vertical line represents the adopted
    3$\sigma$ cut above which  the corresponding time intervals have been discarded. }
\label{fig:bkg}
\end{figure}
In  Fig. \ref{fig:bkg}  an example of the application of this method to  Field  \#6 is shown. Once  the high energy flares were removed, the  0.3--10 keV  background counts distribution was  processed, with  the same 3$\sigma$ clipping method,   in order to remove times during which low energy particle flares were important. These flares are not easily  detected in the 12--14 keV band.  As a result of this selection process the average  time lost due to  particle flares was $<$20$\%$ and 2 observations  were completely lost \citep[see][]{has06}. \\
 An important feature observable  in the background spectrum of both MOS and PN  CCDs   is the Al K$\alpha$ (1.48 keV) fluorescent emission. In the PN background   two strong Cu lines are also present at $\sim$ 7.4 keV and $\sim$ 8.0 keV.  Since these emission lines could affect the scientific results,   the 7.2--7.6 keV and 7.8--8.2 keV energy bands in the PN and the 1.45--1.54 keV band (in PN and MOS) were excluded from the detectors events. 
Images were then created in the  0.5--2 keV, 2--4.5 keV and 4.5--10 keV energy
bands with a pixel size of 4 arcsec.   Single and double  events were used to
construct the  PN images, while MOS images were created using all  valid event
patterns. Out-of-Time (OOT) events appear when a photon hits the CCD during
the read-out process in the IMAGING mode. The result is that the x position of
the event/photon is known, while the y position is unknown due to the readout
and shifting of the charges at this time. For this reason artificial OOT event
files were created.  A new y coordinate  is simulated by randomly shifting the
event along the readout  axis and performing the gain and CTI (charge transfer
inefficiency) correction afterwards. For the PN, in  full frame  mode the OOT
events constitute about 6.3$\%$ of the observing time.    Those files were
filtered in the same way as the event files and the produced images were
subtracted from the event images. Images were then added in order to obtain
PN+MOS mosaics. For each instrument and for each observation, spectrally
weighted exposure maps were created using the SAS task \textit{eexpmap},
assuming a power law model   with photon index $\Gamma$=2.0 in the 0.5--2 keV
band and $\Gamma$=1.7 in the 2--4.5 and 4.5--10 keV bands.  
\subsection{Astrometry correction}
\begin{figure}[t]
\center
\includegraphics[angle=270,width=0.45\textwidth]{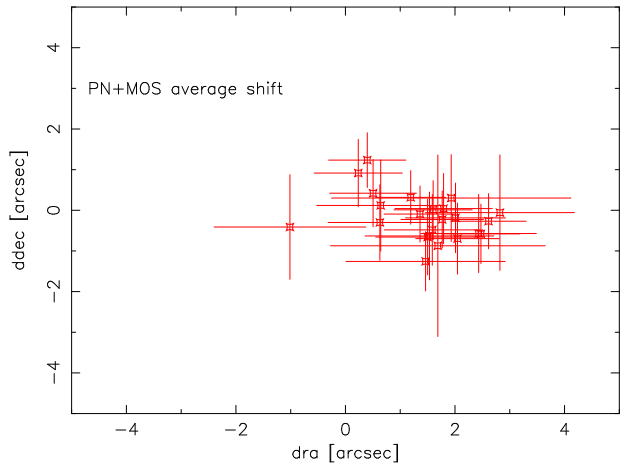}
\caption{Shift between the PN+MOS mosaic and the MEGACAM catalog for each pointing of the XMM-\textit{Newton} COSMOS field. }
\label{fig:astro}
\end{figure}
\begin{figure*}[t]
\begin{center}
\includegraphics[scale=0.8]{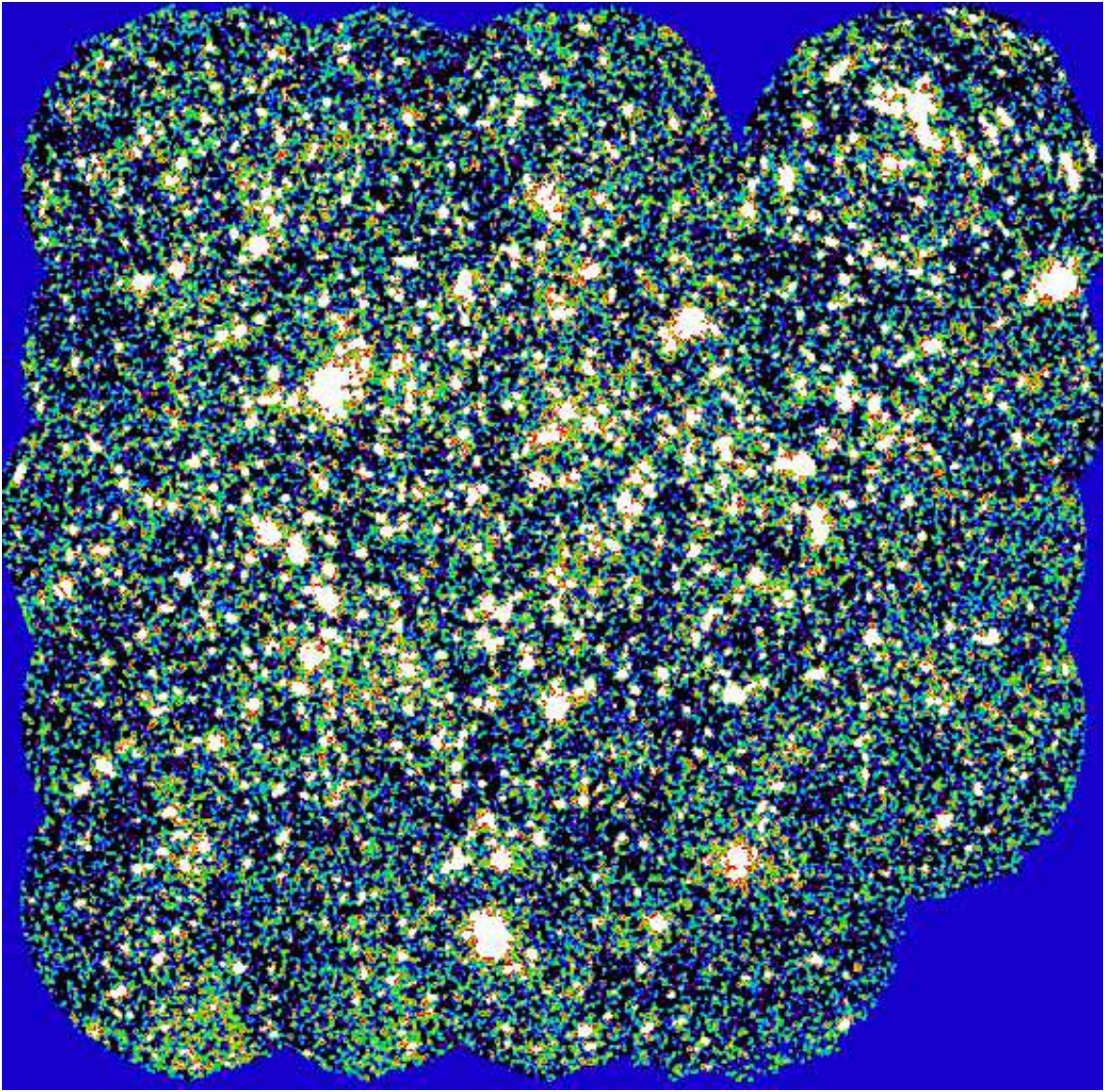}
\caption{ Signal--to--noise ratio map in the 0.5--2 keV band of the XMM--{\it Newton} raster scan in the 
COSMOS field.  The stretch of the color map corresponds to [-0.1$<$S/N$<$1 per pixel]. The scale has been
chosen to enhance the SNR contrasts. If $S$ is the raw  (sources + background) image and $B$ is the model background image, then  the SNR map is obtained by
$SNR=\frac{S-B}{\sqrt{S}}$.  The image
was smoothed with a gaussian filter with $\sigma$=2 pixels. 
Negative values are places where the measured background is smaller than the model background. }       
\label{fig:snr}
\end{center}
\end{figure*}
 In order to correct the astrometry of our \xmm~ observations for each pointing and for each instrument, the 
 produced X-ray  source list  (see next section) was compared with the MEGACAM
 catalog of the COSMOS field \citep{mac06} including all the sources with I
 magnitudes in the range 18-23.   In order to find  the shift between the two
 catalogues,   an optical-X-ray positional correlation was computed using the
 likelihood algorithm included in the SAS task \textit{eposcorr}.  This
 task uses in a purely statistical way all possible counterparts of an X-ray source in the field
 to determine the most likely coordinate displacement. This method is independent of actual spectroscopic identifications, but all post-facto checks using, for example, secure spectroscopic identifications,   have demonstrated its reliability and accuracy.
 Using the magnitude range mentioned above, systematic effects introduced by bright stars and faint background objects are minimized. In the majority of the observations the shift between the three cameras  turned out to be  $<$ 1" (i.e. much smaller than the pixel size of the images used here ). Since the shift between the EPIC cameras is negligible, a correlation between the joint  MOS+PN source list and the optical catalog was calculated to derive the astrometric correction.  For the 23 pointings  presented here, the  shifts between the optical and X-ray catalog are never larger than 3$"$, with  an average shift of the order of   $\Delta \alpha \sim$1.4$"$ and $\Delta \delta \sim$-0.17$"$. The average displacement in the two coordinates
between the PN+MOS mosaic  X-ray positions and the MEGACAM catalog sources for each pointing of the XMM-COSMOS field is shown in  Fig \ref{fig:astro}. The appropriate offset was applied to the event file of each pointing and images and exposure maps were then reproduced with the corrected astrometry.   
 \section  {EPIC source detection}
\begin{deluxetable*}{lcccccc}[!t]
\footnotesize
\tablecaption{Summary of source detection\label{tab:tab1}}
\tablewidth{0pt}
\tablehead{
\colhead{Energy Band} & 
\colhead{$\Gamma$\tablenotemark{a}} & 
\colhead{ECF\tablenotemark{b}} & 
\colhead{S$_{lim}$\tablenotemark{c}} & 
\colhead{N sources\tablenotemark{d}} & 
\colhead{Single detections\tablenotemark{e}}
\\
\colhead{keV}  & 
\colhead{} & 
\colhead{cts s$^{-1}$/10$^{-11}$ erg cm$^{-2}$ s$^{-1}$}  & 
\colhead{\flux~} &  
\colhead{All} & 
\colhead{Point-like} 
}
\startdata
0.5--2   & 2.0 &  10.45 &  7.2$\times$10$^{-16}$ & 1307 & 1281 & 661\\ 
2--4.5    & 1.7 &   1.52 &  4.0$\times$10$^{-15}$ & 735 & 724 &  89\\ 
4.5--10    & 1.7 &   1.21 &  9.7$\times$10$^{-15}$ & 187 & 186 &  3 
\enddata
\tablenotetext{a}{$\Gamma$ is the average spectral index
  assumed in each band.}
 \tablenotetext{b}{The energy conversion factor.Note that the ECF values
  in the second and third rows are the conversion factors from  flux in the
  2--10 keV and 5--10 keV bands to count rate in the 2--4.5 keV  and 4.5--10
  keV bands, respectively. The ECFs are computed by assuming as a mean spectrum an absorbed
   power-law with N$_{H}$=2.6$\times$10$^{20}$ cm$^{-2}$ and  spectral index 
   $\Gamma$=2.0 in the 0.5--2 keV band and  $\Gamma$=1.7 in  the 2--4.5 keV and 4.5--10 keV 
    bands.}
\tablenotetext{c}{The flux
  of the faintest source.} 
\tablenotetext{d}{The total number of sources detected for the
  entire sample and the point-like sample only.} 
  \tablenotetext{e}{the number of sources detected only in one band.}
\end{deluxetable*}

\subsection{Background modeling}

In order to perform the source detection a sophisticated background modeling has been developed. In X-ray observations the background is mainly due to two components, one generated by  undetected faint  sources contributing to the cosmic X-ray background and one arising from soft protons trapped by the terrestrial magnetic field. 
 For this reason two background templates were computed for each instrument and for each pointing,
one for the sky (vignetted) background  \citep{lum02} and one for instrumental and particles
background (unvignetted). To calculate the
normalizations of each template of every pointing, we first performed a wavelet source detection \citep[see][]{fin06}
without sophisticated background subtraction, then we excised the areas
of the detector  where a significant signal due to sources was detected.  The residual area
 is split into two parts depending on  the value of the  effective exposure (i.e. higher and lower than the median value). Using the two templates, we
calculate the coefficients of a system of two linear equations
from which we obtain the normalizations of both:
\begin{eqnarray}
AM^{1}_{v} + BM^{1}_{unv}=C^{1} \\
AM^{2}_{v} + BM^{2}_{unv}=C^{2} 
\end{eqnarray}
where A and B are the normalization factors, M$^{1,2}_{v}$ the vignetted templates in the region with  effective exposure higher and lower than the median , M$^{1,2}_{unv}$ the unvignetted templates and C$^{1,2}$  are   the background counts in the two regions.  The region with effective exposure lower than
the median (i.e. high vignetting, $\gtrsim$7' off-axis) is dominated by the instrumental background, while the region with higher effective exposure   is dominated by the sky background. Therefore with this method we have the advantage of better fitting the two components of the background. The standard method for estimating the background, based on the spline functions  used in the ~\xmm~ pipelines, returned in our case significant residuals.
The excellent result of this technique can be seen in the signal-to-noise (SNR) map in Fig \ref{fig:snr}: despite the 
significant variations in exposure time and average background level from
pointing to pointing, a rather homogeneous signal--to--noise ratio is 
achieved across the whole mosaic.  It is worth noting that also pixels
  with  negative values  are shown in the map; these are 
located where the background model is higher than the measured background.
\subsection{Maximum likelihood detection}
In each pointing the source detection was conducted on the combined images of the different instruments in the three energy bands mentioned above using the SAS tasks \textit{eboxdetect} and \textit{emldetect}. 
As a first step, the sliding cell detection algorithm \textit{eboxdetect} was run on the images in the three energy bands. In this procedure source counts were collected in cells of 5$\times$5 pixels adopting a  low threshold in the detection likelihood (i.e. likemin=4). The source list produced by \textit{eboxdetect} was then  used as input for \textit{emldetect}. For all the sources detected with the sliding cell method this task performs a maximum likelihood PSF fit. In this way refined positions and fluxes  for the sources  were determined.  Due to the particular pattern of our observations \citep[see][]{has06}, the same source could be detected in up to 4 different pointings. For this reason both \textit{eboxdetect} and \textit{emldetect} were  run in raster  mode. The source parameters (position and flux) were fitted simultaneously on all the observations where the source is observable, taking into account the PSF at the source position in each pointing. As likelihood threshold  for  the detection, we adopted the value det\_ml=6.
 This parameter is related to the probability of a random Poissonian fluctuation having caused the observed source counts:
\begin{equation}
det\_ml=-lnP_{random}
\end{equation}
In principle, the expected number of spurious sources could be estimated as the product of   the probability for a random  Poisson  fluctuation exceeding the likelihood threshold times the number of statistically independent trials, N$_{trial}$. 
For  a simple box detection algorithm  N$_{trial}$ would be approximately given by the number of independent source detection cells across the field of view. For the complex multi-stage source detection algorithm,
 like the one applied here, N$_{trial}$ cannot be calculated analytically, but has to be estimated through Monte Carlo simulations.   These simulations, which are discussed in  Section 4, return  a number of spurious sources
 of $\sim$2\% at the likelihood level chosen.
  All the sources were fitted  with a PSF template convolved with a  beta model \citep{cav}. Sources which have a core radius significantly larger than the PSF are flagged as extended ({\it ext} parameter $>0$). 
 
A total of 1307, 735 and 187 X-ray sources were detected in the three bands. 
Of these sources, twenty-six were classified
 as extended. The analysis of the X-ray extended sources in the COSMOS field
 is beyond the scope of this work; these sources are extensively discussed by
 \cite{fin06}. 
 A total of 1281, 724 and 186 point-like sources were detected in the three
 bands down to  limiting fluxes of 7.2$\times$10$^{-16}$ erg cm$^{-2}$
 s$^{-1}$, 4.7$\times$10$^{-15}$ erg cm$^{-2}$ s$^{-1}$ and
 9.7$\times$10$^{-15}$ erg cm$^{-2}$ s$^{-1}$ respectively. The minimum number of net counts
for the detected  sources is $\sim$21, 17 and 27  in the three bands,
 respectively. A total of 1390 independent point-like sources have been detected by
 summing the number of sources detected in each band but not in any softer
 energy band.  
 The number of sources detected only in the 0.5--2 keV, 2--4.5 keV, 4.5--10 keV  bands are 661, 89 and 3, respectively.

   From the count rates in the 0.5--2 keV, 2--4.5 keV and 4.5--10 keV  bands the fluxes were obtained  in the  0.5--2 keV, 2--10 keV and 5--10 keV bands respectively using the energy conversion factors (ECF) listed in Table \ref{tab:tab1}, together with a summary of the source detection. The ECF values have been computed using the 
   most recent EPIC response matrices  in the corresponding  energy ranges. As a model, we assumed  power-law spectra with N$_{H}$ =2.6$\times$10$^{20}$ cm$^{-2}$, (corresponding to the average value of N$_{H}$ over the whole COSMOS field \citep{dick90}) and  the same spectral indices used to compute the exposure maps without considering any intrinsic absorption (see Table \ref{tab:tab1}). 
   It is worth noting that the spectral indices and the absorptions of the individual sources can be significantly different from the average values assumed here. In particular \cite{mai06} found that the spectral  indices $\Gamma$  of the XMM-COSMOS sources are in the range 1.5$\div$2.5,  in the CDFS \citet{toz06} measured an
   average photon index   $<\Gamma>\sim$1.75 and similar values were obtained by \citet{kim04} in the CHAMP survey.
    The mean spectrum assumed here is therefore consistent with the  values measured up to now. 
    By  changing the  spectral index of $\pm{0.3}$ the ECFs change of 2\%, 12\%  and 4\% in the 0.5--2 keV, 2--4.5 keV and 4.5--10 keV, respectively.
\begin{figure}[!t]
\begin{center}
\includegraphics[angle=270,width=.45\textwidth]{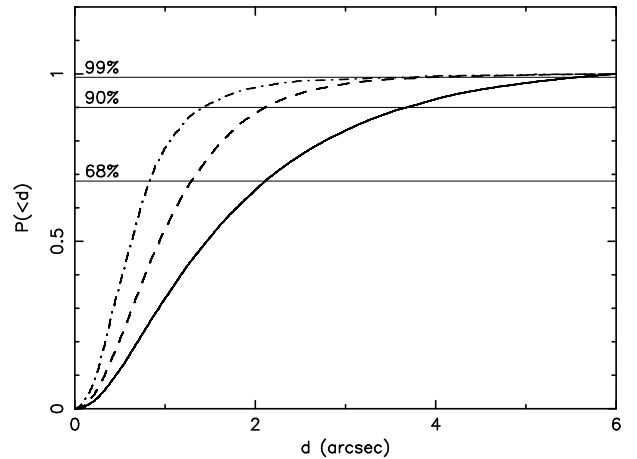}
\caption{ \label{fig:pos}  The cumulative probability to detect a true  source with det$\_$ml$>$6 in a circle of a given radius in the 0.5--2 ($continuous~line$), 2--4.5 ($dashed~line$) and 4.5--10 keV ($dashed-dotted~line$) energy bands. The 68\%, 90\% and 99\%  levels are plotted as horizontal lines. }

\end{center}

\end{figure}

\section{Monte Carlo simulations}
 In order to properly estimate the source detection efficiency and biases,  detailed Monte Carlo simulations were performed \citep[see, e.g.,][]{has93,loa05}. Twenty series of 23 \xmm~  images   were created with the same pattern, exposure maps and background levels as the real data. The PSF of the simulated sources was constructed from the templates available in the  \xmm~ calibration database.   
 The sources were randomly placed in the field of view  according to a
 standard 0.5-2 keV logN-logS distribution \citep{has05}. This was then converted to a 2--4.5 keV and 4.5--10 keV logN-logS assuming that all the sources have the same intrinsic spectrum (a power-law with spectral index $\Gamma=1.7$). 
We then applied, to the simulated fields,  the same source detection procedure used in  the real data.  
 \citet{sch} showed that with the threshold adopted here for source detection,
 which corresponds roughly to the Gaussian 4.5-5$\sigma$, the distortion of
 the slope of the logN-logS due to Poissonian noise is  $<$3$\%$ for a wide
 range of slopes. Therefore, the uncertainties introduced by using a single logN-logS as base for the simulations are negligible.
 A total of 30626, 13579 and 3172 simulated sources were detected  in the 0.5--2 keV, 2--4.5 keV and 4.5--10 keV bands down to  the same  limiting fluxes of the observations.
For every possible pair of input-output sources 
we computed  the quantity
\begin{equation}
R^{2}=\left(\frac{x-x_{0}}{\sigma_{x}}\right)^{2}+\left(\frac{y-y_{0}}{\sigma_{y}}\right)^{2}+\left(\frac{S-S_{0}}{\sigma_{S}}\right)^{2},
\end{equation}
where $x,y$ and $S$ are the position and flux of the detected source and $x_{0},y_{0}$ and $S_{0}$ are the corresponding values for all the simulated sources
We then flag as the most likely associations those with the  minimum value of $R^{2}$. 
The distribution of the positional offsets is plotted in Fig. \ref{fig:pos} for each energy band analyzed. 

\begin{figure}[!ht]
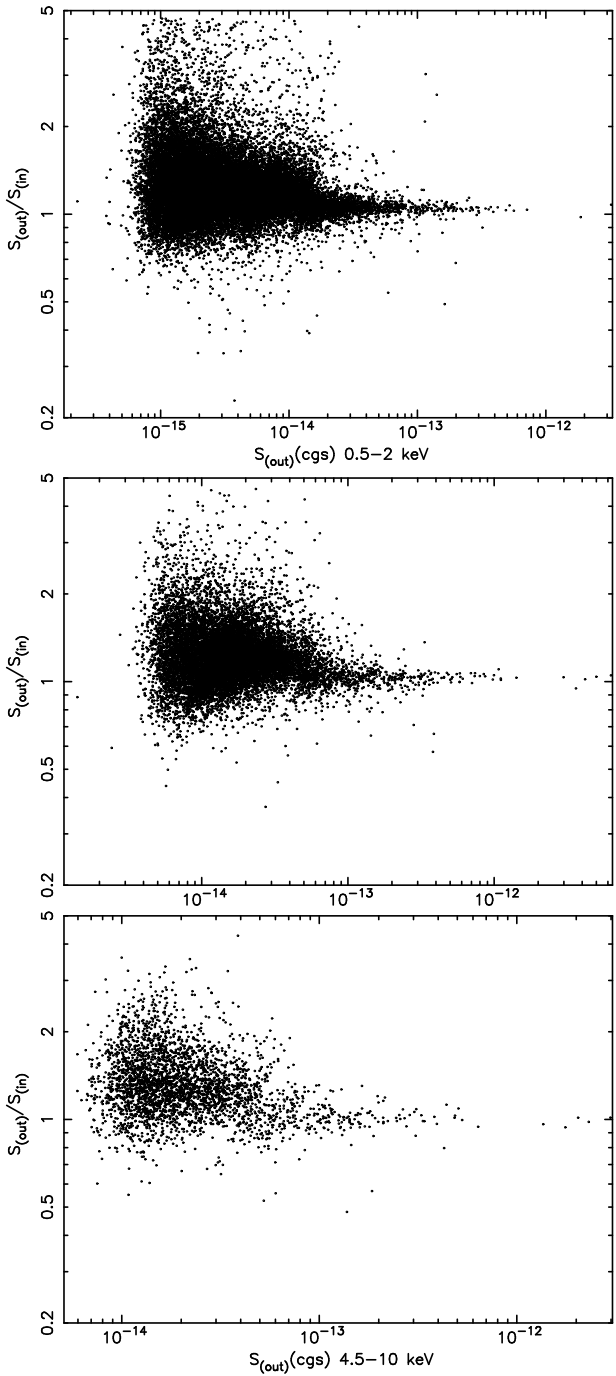

\begin{center}
\includegraphics[angle=270,width=0.45\textwidth]{f6a.eps}
\includegraphics[angle=270,width=0.45\textwidth]{f6b.eps}
\includegraphics[angle=270,width=0.45\textwidth]{f6c.eps}
\caption{ \label{fig:phot} The ratio S$_{out}$/S$_{in}$ as a function of the output detected  flux 
 in the 0.5--2 keV, 2--4.5 keV and 4.5--10 keV bands, respectively$Top~middle~and~bottom~panel$.}      
\end{center}

\end{figure}
We find that  68\% of the sources are detected within 2.1$"$, 1.3$"$ and 0.8$"$ in the 0.5--2 keV, 2--4.5 keV and 4.5--10 keV bands, respectively. Since the detection software fits the position of the source using the information 
available for the three bands together, we expect to be able to detect sources
with an accuracy  of the order of, or somewhat better than, that shown in Fig. \ref{fig:pos}. As in \citet{loa05}, we then define a cut-off radius $r_{cut}$ of 6$"$.  Sources with a displacement larger than  $r_{cut}$ from their  input counterpart are classified as spurious.   
These account for 2.7\%, 0.5\%, 0.6\% of the total number of sources in the 0.5--2 keV, 2--4.5 keV and 4.5--10 keV bands, respectively. 
Source confusion occurs when two or more sources fall in a single resolution element of the detector and result as a
single detected source with an amplified flux.  
In order to determine  the influence of the source confusion we adopted  the method described in \citet{has98}. 
We define as $"confused"$ sources those for which $S_{out}/(S_{in}+3*\sigma_{out})> 1.5$ (where $\sigma_{out}$ is the 1$\sigma$ error on the output flux).   The fraction of $"confused"$ sources is 0.8$\%$, 0.15$\%$ and $<$0.1$\%$ in the 0.5--2 keV, 2--4.5 keV and 4.5--10 keV bands, respectively.  \\
The photometry was also tested; the ratio of  output to  input fluxes in the simulation is plotted  in Fig.\ref{fig:phot}.

 At bright fluxes this ratio is consistent with one, while at fainter  fluxes the distribution
of $S_{out}/S_{in}$ becomes wider, mainly because of increasing errors, and
skewed toward values greater than one. This skewness of the distribution can
be explained mainly by two effects: a) source confusion and  b) Eddington Bias
\citep{edd}.  While source  confusion, as defined above, affects only a small
fraction of the sources, the Eddington bias results in a systematic upward offset
of the detected flux. The magnitude of this effect depends on the shape of the
logN-logS distribution and the statistical error on the measured flux.  Since
there are many more faint  than bright sources, uncertainties in the measured
flux will result in more sources being up-scattered than down-scattered.
Together with this, the fact that in the 4.5--10 keV band we are sampling  a
flux  region in which the logN-logS is steeper than  in the other bands (see
Section 5), explains  why such an effect is more evident in the 4.5--10 keV
band. 

\begin{figure}[ht]
\begin{center}
\includegraphics[angle=270,width=0.45\textwidth]{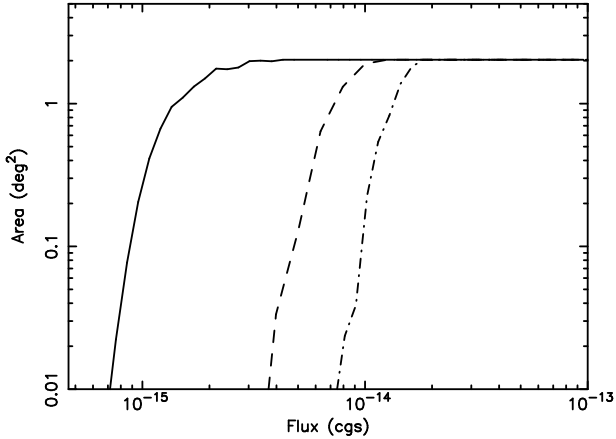}
\caption{ The sky coverage vs. flux relation in the 0.5-2, 2-10 and 5-10 keV energy bands (respectively continuous, dashed and  dash-dotted line), resulting from the simulations described in the text.}     
\end{center}
\label{fig:area}
\end{figure}

Besides assessing the reliability of our source detection procedure, one of the aims of these 
simulations is to provide a precise estimation of the completeness function of our survey, known also as sky coverage.
We constructed our sky-coverage  ($\Omega$) vs. flux relation by dividing the number of detected  sources by the number of input sources as a function of the flux and rescaling it to the sky simulated area. Having analyzed the simulations with the same procedure adopted for the analysis of the data, this method ensures that when computing the source counts distribution (see next section)    all the observational  biases are taken into account and  corrected. The $\Omega$ vs. flux relation  relative to the 0.5--2 keV, 2--10 keV and 5--10 keV bands  is plotted in fig.  \ref{fig:area}.  The total sky area is  2.03 deg$^{2}$  and it is completely observable down to fluxes of $\sim$0.3, 1.3 and 2$\times$10$^{-14}$~\flux\ in the three bands, respectively. The sky coverage  drops to 0 at limiting fluxes of  $\sim$7$\times$10$^{-16}$~\flux, $\sim$4$\times$10$^{-15}$~\flux and  $\sim$9$\times$10$^{-15}$~\flux, in the 0.5--2 keV, 2--10 keV and 5--10 keV bands, respectively.

\section{Source counts}

\begin{deluxetable}{ccc}[!b]
 \footnotesize
\tablecaption{Cumulative number counts\label{key}}
\tablewidth{0pt}
\tablehead{
\colhead{log(S)\tablenotemark{a}} & 
\colhead{$\Omega$\tablenotemark{b}}& 
\colhead{N($>$S)\tablenotemark{c}} 
\\
\colhead{\flux~} & 
\colhead{deg$^{2}$} & 
\colhead{deg$^{-2}$}
 \\
}

\startdata
0.5--2 keV & & \\
\hline
      -13.0 & 2.03 &4.5$\pm{1.5}$ \\
      -13.5 & 2.03 &18.8$\pm{3.1}$ \\
      -14.0 & 2.03 &105.2$\pm{7.0}$ \\
      -14.5 & 2.03 &327.0$\pm{12.7}$ \\
      -15.0 & 0.58 &790$\pm{23.3}$ \\
      -15.1 & 0.12 &931$\pm{53.0}$ \\
\hline
      2--10 keV & &\\
\hline
      -13.0 & 2.03 &8.6$\pm{2.0}$ \\
      -13.5 & 2.03 & 57.0$\pm{5.3}$ \\
      -14.0 & 1.40 &  258.9$\pm{11.6}$ \\
      -14.3 & 0.13 &  600.1$\pm{34.2}$ \\
\hline
      5--10 keV &  &\\
\hline
      -13.5  & 2.03      & 21.3$\pm{3.2}$\\
      -14.0  & 0.35    &      111$\pm{11.0}$ 
\enddata
\tablenotetext{a}{Flux}
\tablenotetext{b}{Sky coverage}
\tablenotetext{b}{Cumulative source counts}
      \end{deluxetable} 
Once the sky coverage is known, the cumulative
source number counts can be easily computed using the following equation:

\begin{equation}
N(>S)=\sum_{i=1}^{\it N_S} \frac{1}{\Omega_{i}} deg^{-2},
\end{equation}
where {\it $N_S$} is the total number of detected sources in the field
with fluxes greater than $S$ and $\Omega_{i}$ is the sky coverage
associated with the flux of the i$^{th}$ source.  The variance of the source number counts is therefore
defined as:
\begin{equation}
\sigma_{i}^{2}=\sum_{i=1}^{\it N_S}\left(\frac{1}{\Omega_{i}}\right)^{2}.
\end{equation} 

\begin{figure}[!t]
\begin{center}
\includegraphics[angle=0,width=.45\textwidth]{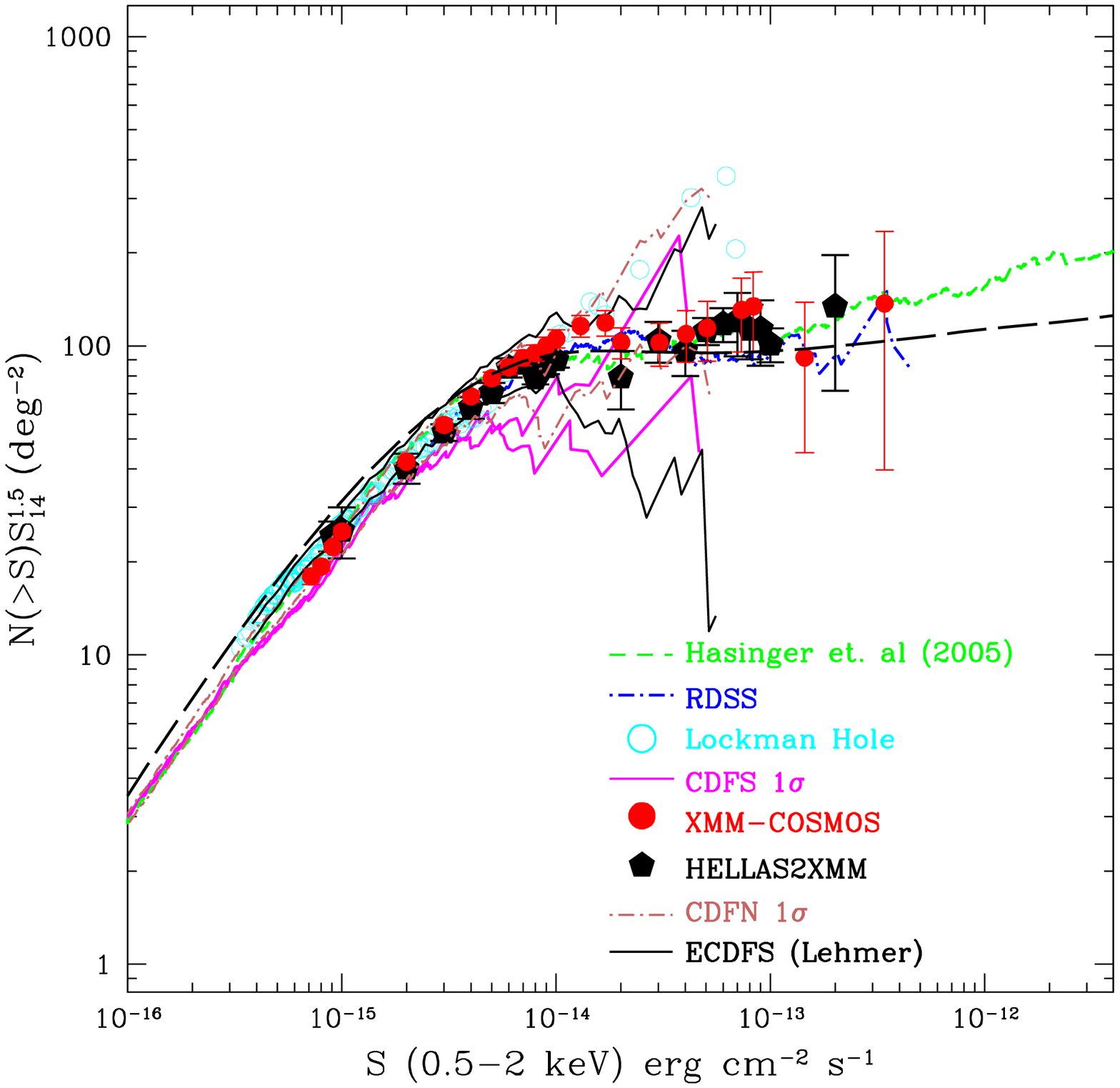}

\caption{The 0.5-2 keV logN-logS of the XMM-COSMOS ($red~dots$) sources compared with the ROSAT medium sensitivity survey \citep{has93} ($blue~dot~dashed~line$),  combined ROSAT, \xmm ~,  {\it Chandra}     \citep{has05} ($green~dashed~line$), the {\it Chandra} deep field south 1$\sigma$ error tie  \citep{ros02} ($magenta~continuous~line$), the {\it Chandra} deep field north 1$\sigma$ error tie  \citep{bau04} ($pink~dot-dashed~line$), the 100 ks of the \xmm ~ Lockman hole ~\citep{has01} ($cyan~circles$), the HELLAS2XMM \citep{bal} ($black~pentagons$)  and the extended
CDFS \citep{leh05} ($black~continuous~line$) surveys.  The overlayed black-dashed line represents the logN-logS predicted by the model of \citet{gil06}.
The source number counts are plotted scaled by S$^{1.5}$ in order to highlight the  deviation from the Euclidean behavior.}
\label{fig:logn1}
\end{center}
\end{figure}

\begin{figure}[!t]
\begin{center}
\includegraphics[angle=0,width=.45\textwidth]{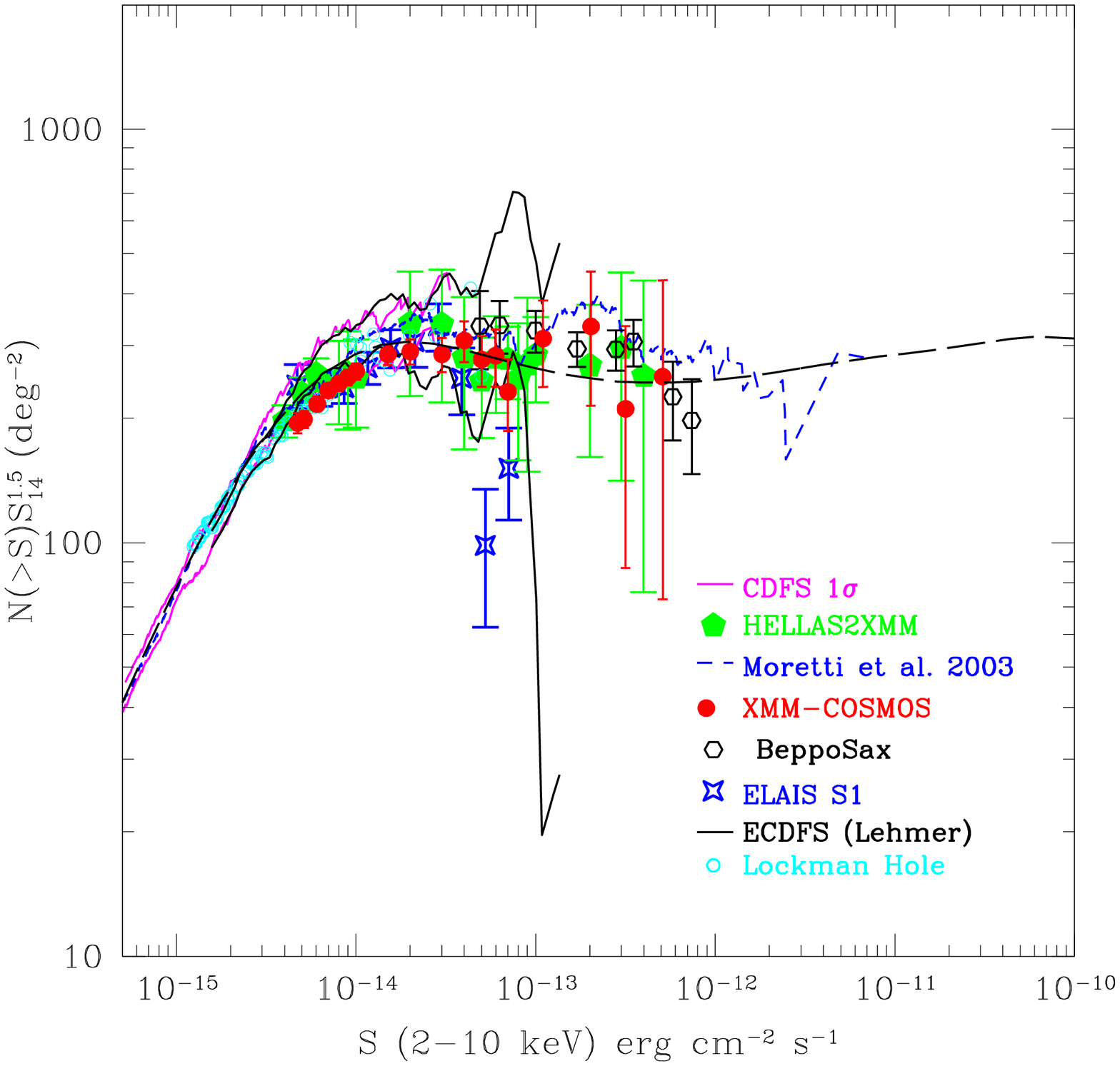}

\caption{ The 2-10 keV logN-logS of  the XMM-COSMOS ($red~dots$) sources compared with     
{\it Chandra},
 \xmm~ and ASCA ($blue~dashed~line$) \citep{mor03}, HELLAS BeppoSAX \citep{gio00} 
 ($black~hexagons$) the  Chandra deep field south 1$\sigma$ error tie \citep{ros02} ($magenta~continuous~line$),  
 the  HELLAS2XMM ($green~pentagons$) \citep{bal},  the ELAIS S1 ($blue~stars$) \citep{puc06}, the extended
CDFS \citep{leh05} 1$\sigma$ error tie ($black~continuous~line$) and the 100 ks of the Lockman
hole ($cian~open~circles$) \citep{has01}. The overlayed black-dashed line represents the logN-logS predicted by the model of \citet{gil06}. The source number counts are plotted scaled by S$^{1.5}$ in order to highlight the  deviation from the Euclidean behavior.}
\label{fig:logn2}
\end{center}
\end{figure}

\begin{figure}[!t]
\begin{center}
\includegraphics[angle=0,width=.45\textwidth]{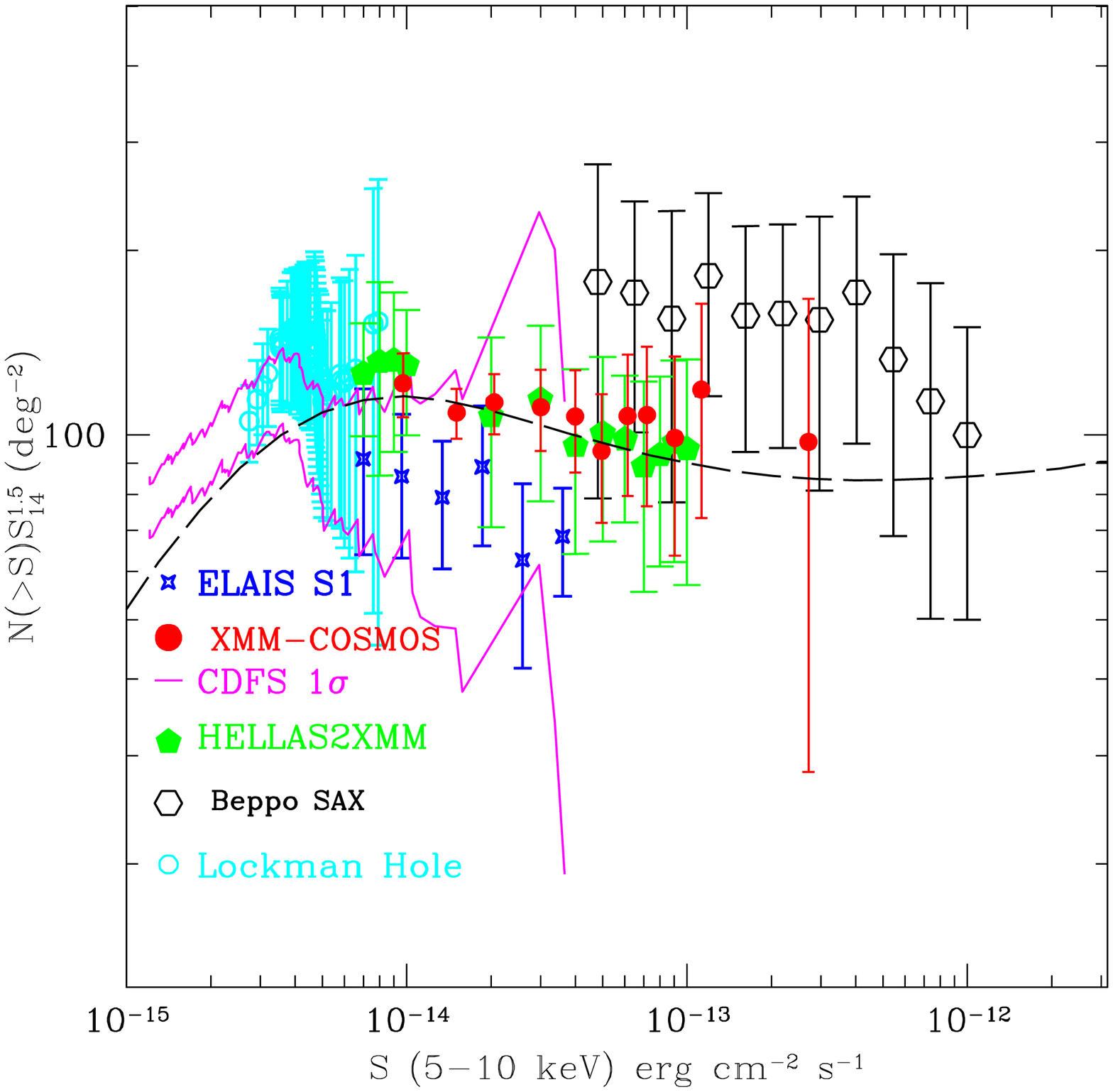}
\caption{ The 5-10 keV logN-logS of the XMM-COSMOS ($red~dots$) sources compared with the HELLAS2XMM \citep{bal} ($green~pentagons$), the Chandra  deep field south 1$\sigma$ error tie \citep{ros02}  ($magenta~continuous~line$) , the HELLAS-BeppoSAX data from \cite{fio01} ($black~hexagons$),  the ELAIS S1 ($blue~stars$) \citep{puc06} and the 100 ks of the Lockman Hole ($cyan~open~circles$) \citep{has01}. The overlayed black-dashed line represents the logN-logS predicted by the model of \citet{gil06}. The source number counts are plotted scaled by S$^{1.5}$ in order to highlight  deviation from the Euclidean behavior.}

\label{fig:pippo}
\end{center}
\end{figure}

Source number counts are reported in Table \ref{key}. 
The cumulative number counts, normalized to the Euclidean slope (multiplied by
S$^{1.5}$), are shown in 
Figures \ref{fig:logn1}, \ref{fig:logn2} and \ref{fig:pippo}, 

for the 0.5--2 keV, 2--10 keV and 5--10 keV energy ranges, respectively.
With such a representation, the deviations from the
Euclidean slope are clearly evident as well as the 
flattening of the counts towards faint fluxes. 
Source counts are compared with the findings of other deep and shallow surveys
collected from the literature. The plotted reference results  were selected
 in order to sample a flux range  as wide as possible and at the same 
 time to keep the plots   as clear as  possible.
As discussed in the previous section, the sky coverage $\Omega$ was derived from realistic 
Monte Carlo simulations and therefore no 
further correction for the Eddington Bias is required. 

 In order to parameterize our relations, 
we performed a maximum likelihood fit  to the unbinned differential counts.
We assumed a broken power-law  model for the 0.5--2 keV and 2--10 keV bands:

\begin{equation}
n(S)=\frac{dN}{ds}= \left\{\begin{array}{ll}
A\,S^{-\alpha_{1}} & S>S_{b} \\
B\,S^{-\alpha_{2}} & S \leq S_{b},\\
\end{array}
\right.
\end{equation}

where $A=B\,S_{b}^{\alpha_{1}-\alpha_{2}}$ is the normalization, 
$\alpha_{1}$ is the bright end slope, $\alpha_{2}$ the faint end slope, 
$S_{b}$  the break flux, and $S$ the flux in units of 10$^{-14}$ \flux~. 
 Notice that using  the maximum likelihood method, the fit is not dependent on the  data binning
and therefore we can make full  use of  the whole dataset. Moreover, the normalization $A$ is not 
a parameter of the fit, but it is obtained by imposing the condition that the number of expected 
sources from the best 
fit model is equal to the  total observed number.
In the 0.5--2 keV energy band
  the best fit parameters are $\alpha_{1}$=2.60$^{+0.15}_{-0.18}$, $\alpha_{2}$=1.65$\pm{0.05}$,
$S_{b}$=1.55$^{+0.28}_{-0.24}$ $\times$10$^{-14}$ \flux ~ and $A$=123. 
Translating this value of the normalization to that for the cumulative 
distribution at 2$\times$10$^{-15}$ \flux ~, which is usually used in the literature
for {\it Chandra} surveys, we obtain $A_{15}\sim$450 which is fully consistent with 
most of previous works where a fit result is presented \citep{has93,mus00,has01,bal,ros02,bau04,kim04,has05,ken05}, but
significantly lower than that found in the CLASXS survey \citep{ya04}.  
In the 2--10 keV band the best fit parameters are $\alpha_{1}$=2.43$\pm{0.10}$, $\alpha_{2}$=1.59$\pm{0.33}$,
$S_{b}$=1.02$^{+0.25}_{-0.19}$ $\times$10$^{-14}$ \flux ~ and $A$=266. The latest value translates into  
$A_{15}\sim$1250. Also in this band, our results are in  agreement  with previous surveys within 1$\sigma$, with
the exception of the CLASXS survey  which is $\sim$30\% higher in this band  \citep{ya04}.
In the 5--10 keV energy bands, where
the differential counts do not show any  evidence for a break in the sampled 
flux range, we assumed a single power-law model of the form:

 \begin{equation}
n(S)=\frac{dN}{ds}=A\,S^{-\alpha_{1}},
 \end{equation}

for which the best fit parameters are found to be $A$=102 and 
$\alpha_{1}$=2.36$\pm{0.1}$. 

In the soft 0.5-2 keV (Fig.~8) energy range  a visual inspection 
of the various datasets  suggests a remarkably good 
agreement between XMM-COSMOS and literature data\footnote{
In particular, it is worthwhile to notice the good agreement between the XMM-COSMOS and the \citet{has05} logN-logS,
which has been used as input in our simulations. This good agreement can be considered as an a posteriori support
of the reliability of those results from the simulations which depend on the assumed  input logN-logS.}.
\par\noindent
In the 2--10 keV band the XMM-COSMOS counts bridge the gap between deep
field observations \citep{ros02} and shallower large area BeppoSAX 
\citep{gio00} and XMM--{\it Newton} surveys \citep{bal}.
At relatively bright fluxes ($> 10^{-14}$ \flux) the XMM-COSMOS logN--logS 
nicely matches previous measurements, though providing a much more robust
estimate of the source counts thanks to the  much smaller statistical 
errors. 
\par\noindent
A major step forward in the determination of X--ray source counts
is achieved in the  5--10 keV band, where the previously existing data from
different surveys show very significant differences.
Thanks to XMM--COSMOS, a solid measure of the hard-X  logN--logS 
in the flux interval 10$^{-14}$ -- 10$^{-13}$  \flux~ is obtained for the
first time. From Fig. \ref{fig:pippo} we notice that the normalization 
of the XMM--COSMOS logN--logS  is  slightly higher than ($\sim$10\%), although
consistent at 1$\sigma$ with, that measured by 
{\it Chandra} while, ELAIS S1 \citep{puc06} source counts are 30\% lower.    However, in the 
overlapping flux range the latter is characterized by large errors due to the
small number of relatively bright sources in the {\it Chandra} deep fields.
Interestingly  enough, the XMM--COSMOS counts match nicely, with smaller errors, 
those of the wide area HELLAS2XMM survey (Baldi et al. 2002), while the pioneering 
measurements of BeppoSAX \citep{fio01} are systematically  higher than the counts from XMM-COSMOS.   

\subsection{Resolved fraction of the X-ray background}
 One of the main aims of XMM-COSMOS with its large and medium--deep coverage,
is to provide a solid census  of the X-ray source population to be compared 
with  observations and  models of  AGN evolution.
According to recent synthesis models  
\citep[see e.g.;][]{com95,gil06,wor}  a high fraction 
of heavily obscured AGN is necessary to explain the spectral shape
and the intensity  of the X-ray background (XRB).
We examine therefore which fraction of XRB  is resolved into discrete sources
in our survey. 

As a first test, we computed the flux which XMM-COSMOS itself resolves into discrete sources 
 by summing their fluxes weighted on the sky coverage in the 0.5--2 keV, 2--10 keV  and  5-10 keV 
energy bands.   
As in \cite{wor} we used as reference value of the normalization at 1 keV of the XRB spectrum 
that of \cite{del} which assumes that the spectral
shape  in the 1-10 keV band is a power-law with spectral index $\Gamma$=1.4 and a normalization 
at 1keV  of 
11.6 keV cm$^{-2}$ s$^{-1}$ keV$^{-1}$.  The latter value corresponds to a flux of 0.80, 2.31 and 1.27$\times10^{-11}$ \xrb~ in the 0.5--2 keV, 2--10 keV  and  5-10 keV 
energy bands, respectively.
In the 0.5--2 keV band we measure a  contribution of the sources to the XRB which corresponds to
 a normalization
at 1 keV of 
0.49$\pm{0.08}\times10^{-11}$ \xrb. The corresponding values in the 
2--10 keV and 5--10 keV bands are  0.92$\pm{0.22}\times10^{-11}$ \xrb~  and 
  0.28$\pm{0.15}\times10^{-11}$ \xrb. 
Therefore XMM-COSMOS resolves by itself $\sim$65$\%$, $\sim$40$\%$ and $\sim$22$\%$ of the XRB into
discrete sources in the  0.5--2 keV, 2--10 keV and 5--10 keV energy bands, respectively. 
It is worth noticing that the flux measured by \citet{del} is the  highest measured in literature
in the 1--10 keV energy range \citep[see e.g.][for a complete collection]{gil06}.   
    It is also worth noticing that we computed the fraction of resolved XRB 
     by assuming that all the sources have the same spectrum. Therefore, in our estimate, 
 effects due to the broad absorption and spectral index distributions of AGNs are not included.

In Figs. \ref{fig:logn1}, \ref{fig:logn2} and \ref{fig:pippo} we compared our logN-logS
 to those predicted by the recent XRB model of \citet{gil06}. 
This model makes use of the most recent observational constraints on the AGN populations 
 and includes  
a conspicuous
fraction of Compton thick AGN which, however, are not expected to significantly  contribute to the 
XMM-COSMOS counts. In the 0.5--2 keV band a direct comparison of our data with the model shows 
a 1$\sigma$
 agreement at the bright end.  At the faint end the model predicts a slightly higher
normalization when compared to most of the plotted data, including ours. 
A similar behavior is observed in the 2--10 keV band. It is worth noticing that in the model 
the average unabsorbed power-law spectral index of the sources 
is $<\Gamma>\sim$1.8 in the flux interval sampled by XMM-COSMOS (see Fig. 19 in \citet{gil06}).
  Since in our data analysis we assumed $<\Gamma>$=1.7 we expect in this band a slight ($\sim$10\%)
 underestimation of the fluxes when compared to those of the  model. 
 This effect is almost negligible (i.e. $<$5\%) in the other bands investigated here. 
  
  By integrating our best fit  2--10 keV logN-logS between infinite and zero, and assuming that the
  slope of the "real" logN-logS remains constant down to low fluxes, we estimated the total contribution of 
   AGNs to the XRB. We predict a total flux of AGNs in the XRB of 1.25$\times10^{-11}$ \xrb. 
  This value is $\sim$40\% lower than that measured by \citet{del}, and $\sim$ 25\% smaller than those
  obtained by integration the model logN-logS of \citet{gil06} which predicts  a flux of  $\sim$1.66$\times10^{-11}$ \xrb.
   This discrepancy between our predicted flux 
  and that of \citet{gil06}, could arise by the fact that, in our measurement, we consider that all the  sources 
have  the same spectrum and from  statistical uncertainty of the logN-logS parameters. By assuming an average spectral
  index $<\Gamma>$=1.4 for all our sources, we obtain a value for the total flux of the AGNs of $\sim$1.48$\times10^{-11}$ \xrb~ which is consistent, within the statistical uncertanties, with the prediction of the model.  Considering  the 
  total flux of the XRB predicted by the model and our estimate
  from the logN-logS distributions, in the 2--10 keV band, 
  XMM-COSMOS resolves the $\sim$55--65\% of the total flux of the XRB into discrete sources.

It is interesting to observe   how in the 5--10 keV band our data are in good 
agreement with the prediction of the model.  
This  result is particularly important since in this band it is expected the major contribution
 of highly absorbed AGN, which are an important ingredient of  the XRB models. 
 A detailed analysis of the spectral properties of the brightest X-ray 
sources in XMM-COSMOS is presented by \citet{mai06}.

\section{Sample variance}
The amplitude of  source counts distributions  varies
significantly among different surveys \citep[see e.g.][and references therein]{ya03,cap05}.
This  "sample variance", can be explained as a combination of Poissonian variations and effects
due to the clustering of sources \citep{pee,ya03}.
The variance of  counts in cells for sources which are angularly correlated can be obtained with:
\begin{equation}
\label{cvar}
<(N-\mathcal{N}\Omega)^{2}>=\mathcal{N}\Omega+\mathcal{N}^{2}\,\int\,d\Omega_{1}\,d\Omega_{2}w(\theta_{1,2})
\end{equation}
 where $\mathcal{N}$ is the mean density of objects in the sky, $\Omega$ is the cell size, 
 and $w(\theta_{1,2})$ is the angular two--point correlation function. 
The first term of Eq. \ref{cvar} is the Poissonian variance and the second
 term is  introduced by the large scale structure.
In order to determine whether  the differences observed in  the source counts of different
 surveys could arise 
from the clustering of X-ray sources, we estimated  the amplitude of the fluctuations from our data, 
 by producing subsamples of our survey with areas  comparable to those of. e.g., {\it Chandra} 
surveys. \\
The XMM-COSMOS field and the Monte Carlo sample fields  of Section 4 were divided in 4,9,16 
and 25 square boxes.
Making use of the 0.5--2 keV energy band data, we  computed for each subfield, the  ratio of
 the number of real
 sources to the number of random  sources. In order to prevent incompleteness artifacts, we
 conservatively cut
 the limiting flux of the random and data sample to 5$\times$10$^{-15}$ \flux. At this flux
  our survey is
 complete over the entire area.  In order to avoid artifacts introduced by the missing pointings in the 
 external part of the field of view,
 we concentrated  our analysis to the central 80$'\times$80$'$.  
\begin{figure}[!t]
\begin{center}
\includegraphics[angle=0,width=.45\textwidth]{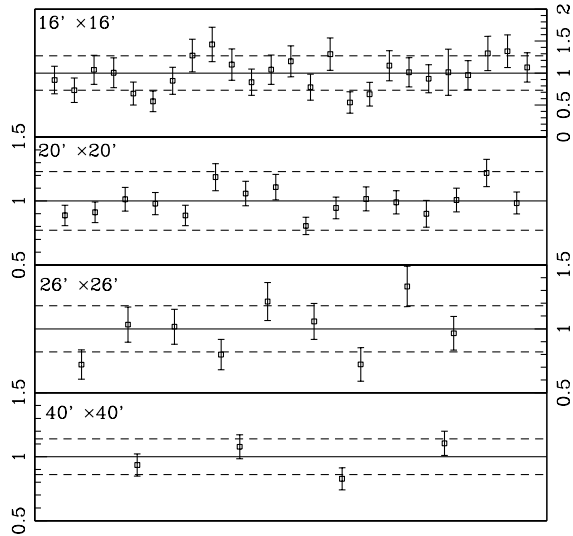}
\caption{  The  counts in cell fluctuations  within the XMM-COSMOS field.
 The  data are normalized to a random distributed field in boxes of 40',26',20' and 16' of side, respectively. The dashed lines represent the 1$\sigma$ expected fluctuation.}

\label{fig:varianza}
\end{center}
\end{figure}
In Fig.\ref{fig:varianza} we plot the ratio of the data to the random sample as a function
 of the size of the
 cells under investigation.  The measured fractional standard deviations  of the sample is reported
 in Table~\ref{tab:fit}.
Using Eq.~\ref{cvar}  we computed the expected amplitude of source counts fluctuations 
 with $w(\theta_{1,2})$
 taken from \cite{myj06}. They computed the X-ray two--point correlation function
 in the XMM-COSMOS field and  
detected clustering signal with angular correlation length  $\theta_{0}$
 $\sim$1.9$"$ $\sim$0.8$"$ and $\sim$6$"$ 
in the  0.5--2 keV, 2--4.5 keV and 4.5--10 keV band, respectively. The observed
 slope is $\gamma$=1.8 in all
 the energy bands. 

The predicted fractional standard deviations are therefore 0.13, 0.19, 0.23 and 0.28
  on scales of 
 0.44 deg$^{2}$, 0.19 deg$^{2}$, 0.11 deg$^{2}$  and 0.07 deg$^{2}$, respectively.  
 These values are in good agreement with those observed in the sub-samples
 of our dataset as shown by the value of the fitted $\chi^{2}$ to the counts in cell fluctuations (see Table \ref{tab:fit}). 
 As shown in Table \ref{tab:fit}, at this limiting flux and on the areas considered
 here the main
 contribution to the source counts fluctuations is from the Poissonian noise. 
  At the flux limit assumed here, the ratio  $\sigma_{cl}$/$\sigma_{p}$  increases from  $\sim$0.5 on the smallest  scale (16 x 16 arcmin)
 to $\sim$0.85 on the largest scale (40 x 40 arcmin). This ratio scales as:
\begin{equation}
\sigma_{cl}/ \sigma_{p}\propto  \mathcal{N}^{0.5} \theta_{0}^{(\gamma-1)/2} a^{(3-\gamma)/2},  
\end{equation}
where $\mathcal{N}$   deg$^{-2}$ is the surface density of the sources, 
 $\theta_{0}$ (deg) is the
 angular correlation length and $a$ (deg) is the size of the cell. 
  In order to estimate at which flux limit fluctuations introduced by the large scale 
 structure are predominant, 
  we estimate that  
 $\sigma_{cl}$/$\sigma_{p}$  would
 be $\sim$1 on the smallest scale, corresponding to a Chandra ACIS field
 of view (16 x 16 arcmin) 
 at a surface density of the order of $\sim$900 deg$^{-2}$, corresponding 
to a 0.5-2 keV flux 
S$\sim$ 8$\times$10$^{-16}$ \flux. 
 At even fainter fluxes the dominant contribution to the
 total expected source counts fluctuations on this area ($\sigma_{exp}$ ) comes from the
 large scale structure, therefore the contribution of statistical fluctuations
 becomes less important. 
With the same procedure, we can estimate the total expected fluctuations ($\sigma_{exp}$)
 and the relative importance of $\sigma_{p}$  and  $\sigma_{cl}$ also for 
the hard band (5-10 keV), 
even if in this band we do not have enough statistics to divide our field in sub-samples. 
Using the formal best fit for $\theta_{0}$ in this band  found by Miyaji et al. (2007)
 ($\theta_{0}$=6$"$), we find that at the faintest 5-10 keV flux (S$\sim$10$^{-14}$ \flux)
 sampled by the XMM-COSMOS survey ($\mathcal{N} \sim$ 110 deg$^{-2}$ ) the ratio 
 $\sigma_{cl}$/$\sigma_{p}$ is smaller than one on all the four scales here analyzed,
 with a total expected standard deviation of the fluctuations ranging from $\sim$0.20
 on the largest scale to $\sim$0.40 on the smallest scale. These values for $\sigma_{exp}$
 are significantly larger than those shown in Table 3 for the soft band, because in the
 hard band the surface density of sources is lower and the angular correlation length
 is higher than in the soft band.

This analysis is at least qualitatively consistent with Figures 8 and 10, which show a 
 significantly larger dispersion in the data from different surveys in the hard band than
 in the soft band. Moreover, the results here discussed are also consistent with the
 observed fluctuations in the deep Chandra fields (see, for example,  Bauer et al. 2004).
 Large area, moderately deep surveys like XMM-COSMOS are needed to overcome the problem 
of low counting statistics, typical of deep pencil beam surveys, and, at the same time, 
to provide a robust estimate of the effect of large scale structure on observed source 
counts.

  As a final consideration, we tried to compute  the expected intrinsic variance of
 XMM-COSMOS. This 
  estimate must be made with care since we have only one sample on this scale.
 Assuming that the
  angular correlation function of \cite{myj06} was universal, the residual
 uncertainties on the source counts
  are estimated to be $<$5-6\% in the 0.5--2 keV and 2--10 keV bands, and 
  of the order of the 10\% in the 5--10 keV band.

\begin{deluxetable}{cccccc}
\footnotesize
\tablecaption{\label{tab:fit}  Summary of the 0.5--2 keV sample variance in the COSMOS field. Prediction and observation at a flux limit S$_{lim}$=5$\times$10$^{-15}$ \flux}
\tablewidth{0pt}
\tablehead{
\colhead{Area\tablenotemark{a}} & 
\colhead{$\sigma_{obs}$\tablenotemark{b}} &
\colhead{$\sigma_{p}$\tablenotemark{c}} &
\colhead{$\sigma_{cl}$\tablenotemark{d}} & 
\colhead{$\sigma_{exp}$\tablenotemark{e}}  & 
\colhead{$\chi^{2}$/d.o.f.\tablenotemark{f}} 
\\
\colhead{arcmin$^{2}$}
&   
& 
& 
& 
\\
}
\startdata
   40$'\times$40$'$ & 0.09$\pm{0.04}$ & 0.10 & 0.09 &0.13 & 4.21/3\\
   26$'\times$26$'$     & 0.20$\pm{0.05}$ & 0.15 & 0.10 & 0.19 & 8.93/8\\
   20$'\times$20$'$    & 0.21$\pm{0.04}$ & 0.20 & 0.11 & 0.23 & 16.63/15\\
   16 $'\times$16$'$& 0.24$\pm{0.02}$ & 0.25 & 0.12 & 0.28 & 25.15/24 
     \enddata
\tablenotetext{a}{Size of the independent cells.} 
\tablenotetext{b}{The observed standard 
       deviation.} 
\tablenotetext{c}{The predicted Poissonian standard deviation  $\sigma_{p}$.} 
\tablenotetext{d}{The predicted standard deviation due to clustering $\sigma_{cl}$.} \tablenotetext{e}{The total predicted 
                standard deviations.} . 
\tablenotetext{f}{Value of the fitted $\chi^{2}$/d.o.f.  }
\end{deluxetable}

\section{Summary}

The data analysis of the first run of observations of the XMM-\textit{Newton} 
COSMOS wide field survey has been presented.   
A total of 1390  point-like sources are detected on a contiguous 
area of about 2 deg$^{2}$  down to 
fluxes of  7.2$\times$10$^{-16}$ \flux,  4.0$\times$10$^{-15}$ \flux 
and 9.7$\times$10$^{-15}$ \flux in the 0.5--2 keV, 2--10 keV and 5--10 keV energy bands, 
respectively.  
The detection procedure was tested through Monte Carlo simulations 
which confirmed the high level of accuracy in the determination of the 
source properties (aperture photometry and positioning) 
and allowed us to keep  statistical biases under control. 

A robust estimate of X--ray source counts at both soft and hard energies, 
obtained thanks to the large number of sources detected in the 
XMM--COSMOS survey, is presented in this paper.
The differential logN--logS was fitted with a broken power-law model in 
the 0.5--2 keV and 2--10 keV energy bands, and with a single power-law model in the 
5--10 keV  energy band.
In the soft 0.5--2 keV band, already extensively covered by 
ROSAT, XMM--{\it Newton} and {\it Chandra} surveys over a range 
of fluxes encompassing those sampled by the COSMOS survey, our results 
are in excellent agreement with previous analysis (see Fig \ref{fig:logn1}),
providing an independent evidence of the validity of 
our data analysis procedure. 

The large number of X--ray sources of the COSMOS survey allowed us 
to constrain with unprecedented accuracy the logN-logS parameters 
in the 2--10 and 5--10 keV energy ranges over a range of fluxes 
which were previously poorly constrained. 
Most importantly, in the  hard 5--10 keV band, we were able to 
fill the gap between the deep {\it Chandra} surveys in the CDFS and
CDFN and shallower large area surveys. 
 The deviations from other surveys, which are, however, less than 30\%,
have been explained in terms of low counting statistics of pencil beam
surveys, and partially by the effect of large scale structure.
The major step forward in the determination of hard X--ray source
counts achieved thanks to the XMM--COSMOS survey will provide 
an important reference point for the study of the AGN demography and evolution
especially with applications to obscured AGN. 
More specifically, the evolutionary properties of the
 obscured AGN can be tightly constrained, 
since they are indeed very sensitive, according to the most recent model
of the X--ray background (Gilli, Comastri \& Hasinger 2006), to the shape 
of the hard X--ray source counts around the break flux, which is precisely 
where the COSMOS data play a key role.  
 In this context, we compared  our results to the most 
recent  predictions of the model by \cite{gil06}, finding a remarkable agreement between data 
and model.  

The second pass of the \xmm~ observations in the COSMOS field (600 ks) has 
already started, and is expected to significantly increase the total number of X--ray sources.
  The results of the full XMM--COSMOS survey 
including the complete (AO3+AO4) source catalogue will be the subject of a future
paper. 
It is anticipated that, when completed, the XMM--COSMOS survey will 
provide a number of X--ray sources 
over a large enough contiguous area and down to moderately deep fluxes 
that it will make possible the study of  AGN evolution and their 
connection with the large scale structure in which they reside 
with  unprecedented detail. 
 The COSMOS field has been granted  1.8 Msec observation with {\em Chandra} 
in its central square degree (C-COSMOS, P.I. Martin Elvis). The joint Chandra and \xmm~
observation will provide an unprecedented lab for AGN physics. 
  
\begin{acknowledgments}
This work is based on observations obtained with \xmm~, an ESA science
mission with instruments and contributions directly
 funded by ESA Member States
and the US (NASA). In Germany, the XMM--{\it Newton} project is supported by the
Bundesministerium f\"ur Bildung und Forschung/Deutsches Zentrum f\"ur 
Luft und
Raumfahrt, the Max--Planck Society, and the Heidenhain--Stiftung. 
In Italy, the XMM-COSMOS project is supported by INAF and MIUR under 
grants PRIN/270/2003 and Cofin-03-02-23 and by ASI under grant ASI/INAF
I/023/05/0. 
In the USA the XMM-COSMOS project is supported by NASA NNG04GG40G and
NNG06GD77G. 
This work is based in part on observations obtained with 
MegaPrime/MegaCam, a joint project of CFHT and CEA/DAPNIA, 
at the Canada-France-Hawaii Telescope
(CFHT), and on data products produced at TERAPIX and the Canadian Astronomy
Data Centre. We kindly acknowledge  the anonymous referee for his/her valuable comments
and suggestions.
We gratefully acknowledge the contributions of the entire COSMOS 
collaboration consisting of more than 70 scientists. 
More information on the COSMOS survey is available at
\verb+http://www.astro.caltech.edu/~cosmos+.
 \end{acknowledgments}
 
\end{document}